\documentclass[aps]{revtex4}

\usepackage{graphicx}

\begin{document}

\title{Variational and perturbative approaches to the confined
hydrogen atom with a moving nucleus}

\author{F. M. Fern\'andez}\email{fernande@quimica.unlp.edu.ar}
\affiliation{INIFTA (UNLP, CCT La Plata--CONICET), Divisi\'on Qu\'imica
Te\'orica \\ Blvd. 113 y 64 S/N,
Sucursal 4, Casilla de Correo 16, 1900 La Plata, Argentina}

\author{N. Aquino}
\affiliation{Departamento de F\'{\i}sica, Universidad Aut\'onoma
Metropolitana-Iztapalapa, Apartado Postal 55-534, 09340 M\'exico D.F., M\'exico}

\author{A. Flores-Riveros}
\affiliation{Instituto de F\'{\i}sica, Benem\'erita Universidad Aut\'onoma de
Puebla, Apartado Postal J-48, 72570 Puebla, Pue., M\'exico}

\begin{abstract}
We calculate the ground--state energy and other physical properties of the
hydrogen atom inside a spherical box with an impenetrable wall. We apply the
variational method and perturbation theory and compare both approximate results.
We show that the total, kinetic and potential energies for the moving--nucleus
model are greater than those for the case in which the nucleus is clamped at the
box center.
\end{abstract}

\maketitle

\section{Introduction}

\label{sec:intro}

Atoms and molecules confined into boxes of different shapes and permeability
proved to be suitable simple models for the study of the effect of the
environment or of high pressure. In the later case the pressure is given by $%
p=-dE/d\Omega $ where $E$ is the energy and $\Omega $ the volume of the
confining box. Those models enable us to simulate the effect of pressure on
several properties of atoms and molecules, such as, for example, their
polarizability\cite{F84}. There is a vast literature on the subject and the
reader may find suitable reviews elsewhere\cite{PV09,SPM09,L-K09,A09,L09}.

In all those cases the nucleus was clamped at some chosen point within the
box. For example, in the case of an impenetrable spherical box we know that
the energy is lowest when the nucleus is clamped at origin and increases as
the nucleus approaches the surface\cite{MRU95}. This behavior comes entirely
from the interaction between the electron and the wall and does not tell us
anything about the effect of the hard surface on the nucleus.

If we have an atom in a real environment we should assume that the nucleus
also interacts with that environment. Therefore, it seems reasonable to add
that interaction explicitly into the model. The simplest one is that in
which both the nucleus and electron are affected exactly in the same way by
the impenetrable spherical surface. In such a case we know that the nucleus
cannot have zero kinetic energy because of its interaction with the surface
and, therefore, the clamped--nucleus approach may not be the most realistic
one\cite{F10a,F10b,F10c}.

It seems more reasonable to assume that a repulsive force on the electron
should be an attractive one on the nucleus and conclude that it is more
realistic to choose different boundary conditions for each particle at the
box surface. However, as a first approximation to more elaborate models it
seems sensible to start with an impenetrable wall for both particles and
compare the results of this model with the clamped--nucleus approach.

\section{The model}

\label{sec:model}

The Hamiltonian operator for a nonrelativistic hydrogen atom is
\begin{eqnarray}
\hat{H} &=&\hat{T}+\hat{V}  \nonumber \\
\hat{T} &=&-\frac{\hbar ^{2}}{2m_{e}}\nabla _{e}^{2}-\frac{\hbar ^{2}}{2m_{n}%
}\nabla _{n}^{2}  \nonumber \\
V(r) &=&-\frac{e^{2}}{4\pi \epsilon _{0}r}  \label{eq:H}
\end{eqnarray}
where $m_{e}$ and $m_{n}$ are the masses of the electron and nucleus located
at $\mathbf{r}_{e}$ and $\mathbf{r}_{n}$ with charges $-e$ and $e$,
respectively, $r=|\mathbf{r|}$, $\mathbf{r}=\mathbf{r}_{e}-\mathbf{r}_{n}$, $%
\epsilon _{0}$ is the vacuum permittivity and $\nabla ^{2}$ denotes the
Laplacian in the coordinates indicated by the subscript.

If the atom is confined to a spherical box of radius $R$ with an
impenetrable wall we can formally write the Hamiltonian operator in atomic
units as
\begin{eqnarray}
\hat{H} &=&\hat{T}+V(r)+U(r_{e},r_{n})  \nonumber \\
\hat{T} &=&-\frac{1}{2}\nabla _{e}^{2}-\frac{1}{2m_{n}}\nabla
_{n}^{2},\;V(r)=-\frac{1}{r}  \nonumber \\
U(r_{e},r_{n}) &=&\left\{
\begin{array}{c}
0 \ \mathrm{if } \ r_{e},\ r_{n}\leq R \\
\infty \ \mathrm { if } \ r_{e},\ r_{n}>R
\end{array}
\right.  \label{eq:H_au}
\end{eqnarray}
where $m_n = 1836.15267261$ and the eigenfunctions should
vanish when either $r_{e}=R$ or $r_{n}=R$.

It is well--known that the virial theorem for such a system is given by\cite
{SPM09,A09,FC87}
\begin{equation}
R\frac{dE}{dR}=3\Omega \frac{dE}{d\Omega }=-2\left\langle \hat{T}%
\right\rangle -\left\langle V\right\rangle  \label{eq:VT}
\end{equation}
where $\Omega $ is the volume of the spherical box. This expression gives us
the pressure on the hydrogen atom $p=-dE/d\Omega $ in terms of the kinetic
and potential energies. When $R\rightarrow \infty $ we recover the virial
theorem for the free hydrogen atom
\begin{equation}
2\left\langle \hat{T}_{r}\right\rangle +\left\langle V\right\rangle =0,\;%
\hat{T}_{r}=-\frac{1}{2m}\nabla ^{2}  \label{eq:VT_R->inf}
\end{equation}
where $\hat{T}_{r}$ is the kinetic energy for the relative motion and $%
m=m_{n}/(m_{n}+1)$ is the reduced mass in atomic units.

\section{Methods of calculation}

\label{sec:methods}

\subsection{The variational method}

\label{subsec:variational}

In order to calculate accurate eigenvalues and eigenfunctions of the
hydrogen in a box we resort to the generalized Hylleraas basis set that
proved useful for three--body Coulomb systems\cite{H64,KRS60,SMKS87,FF93}.
For example, for the $S$ states we choose a trial function of the form
\begin{equation}
\varphi (r_{e},r_{n},r)=\left( 1-\frac{r_{e}}{R}\right) \left( 1-\frac{r_{n}%
}{R}\right)
\sum_{k=1}^{N}c_{k}r_{e}^{n_{k}}r_{n}^{m_{k}}r^{l_{k}}e^{-\alpha
_{k}r_{e}-\beta _{k}r_{n}-\gamma _{k}r}  \label{eq:phi_v_gen}
\end{equation}
with linear $c_{k}$ and nonlinear $\alpha _{k},\beta _{k},\gamma _{k}$
variational parameters. The explicitly correlated character of these
functions ensures accurate energies for the ground and low excited states of
free three--body atomic and molecular species with relatively few terms in
the expansion. The effect of the Hamiltonian operator (\ref{eq:H_au}) on
this variational function is simply given by
\begin{eqnarray}
\hat{H}\varphi &=&-\frac{1}{2}\left( \frac{\partial ^{2}}{\partial r_{e}^{2}}%
+\frac{2}{r_{e}}\frac{\partial }{\partial r_{e}}+\frac{%
r_{e}^{2}-r_{n}^{2}+r^{2}}{r_{e}r}\frac{\partial ^{2}}{\partial
r_{e}\partial r}\right) \varphi  \nonumber \\
&&-\frac{1}{2m_{n}}\left( \frac{\partial ^{2}}{\partial r_{n}^{2}}+\frac{2}{%
r_{n}}\frac{\partial }{\partial r_{n}}+\frac{r_{n}^{2}-r_{e}^{2}+r^{2}}{%
r_{n}r}\frac{\partial ^{2}}{\partial r_{n}\partial r}\right) \varphi
\nonumber \\
&&-\frac{m_{n}+1}{2m_{n}}\left( \frac{\partial ^{2}}{\partial r^{2}}+\frac{2%
}{r}\frac{\partial }{\partial r}\right) \varphi -\frac{1}{r}\varphi
\label{eq:Hd_phi_v}
\end{eqnarray}

A recent pedagogical approach to the confined hydrogen atom with a moving
nucleus proposed the simple ansatz
\begin{equation}
\varphi (r_{e},r_{n})=N(\gamma ,R)\left( 1-\frac{r_{e}}{R}\right) \left( 1-%
\frac{r_{n}}{R}\right) e^{-\gamma r}
\end{equation}
where $\gamma $ is a variational parameter and $N(\gamma ,R)$ the
appropriate normalization factor\cite{F10a,F10b}. Since this function yields
the correct result when $R\rightarrow \infty $ then it provides accurate
ground--state energies for large and moderate values of $R$. In this paper
we propose the straightforward generalization
\begin{equation}
\varphi (r_{e},r_{n})=\left( 1-\frac{r_{e}}{R}\right) \left( 1-\frac{r_{n}}{R%
}\right) e^{-\alpha r_{e}-\beta r_{n}-\gamma r}
\sum_{k=1}^{N}c_{k}r_{e}^{n_{k}}r_{n}^{m_{k}}r^{l_{k}}
\label{eq:phi_v_gen_2}
\end{equation}
that may be suitable for the ground and excited $S$ states.

Notice that the cut--off function $(1-r_{e}/R)(1-r_{n}/R)$ tends to unity as
$R\rightarrow \infty $ so that the resulting asatz (\ref{eq:phi_v_gen_2})
will be suitable for the free system. We also expect that under such
condition $\alpha $ and $\beta $ vanish as well as the coefficients $c_{k}$
of those terms with $n_{k}\neq 0$ and $m_{k}\neq 0$.

For comparison we will also consider the hydrogen atom with the nucleus
clamped at origin. A suitable trial function is
\begin{equation}
\varphi ^{CN}(r)=\left( 1-\frac{r}{R}\right) e^{-\delta
r}\sum_{j=0}^{N}d_{j}r^{j}  \label{eq:phi_v_CN}
\end{equation}
where $\delta $ and $d_{j}$ are nonlinear and linear variational parameters,
respectively. This trial function yields the exact $S$ states of the free
hydrogen atom when $R\rightarrow \infty $. For brevity we refer to the
moving--nucleus case (MNC) and clamped--nucleus case (CNC) from now on.

\subsection{Perturbation theory}

\label{subsec:PT}

In the strong--coupling regime $R\rightarrow 0$ the Coulomb interaction is
entirely dominated by the kinetic energy and the problem is almost separable%
\cite{F10a,F10b,F10c}. In the first approximation we have two particles
moving independently within the box and we can apply
Rayleigh--Schr\"{o}dinger perturbation theory by splitting the Hamiltonian
operator into the zero--order or reference
\begin{equation}
\hat{H}_{0}=-\frac{1}{2}\nabla _{e}^{2}-\frac{1}{2m_{n}}\nabla
_{n}^{2}+U(r_{e},r_{n})  \label{eq:H0}
\end{equation}
and perturbation $\hat{H}^{\prime }=-1/r$ parts.

The unperturbed eigenfunctions are given by
\begin{equation}
\psi _{nlmn^{\prime }l^{\prime }m^{\prime }}^{(0)}(\mathbf{r}_{e},\mathbf{r}%
_{n})=\chi _{nlm}(\mathbf{r}_{e})\chi _{n^{\prime }l^{\prime }m^{\prime }}(%
\mathbf{r}_{n})  \label{eq:phi_PT_0}
\end{equation}
were
\begin{equation}
\chi _{nlm}(\mathbf{r})=R_{nl}(r)Y_{lm}(\theta ,\phi )
\end{equation}
and
\begin{equation}
R_{nl}(r)=N_{n,l}j_{l}(x_{l,n}r/R)
\end{equation}
In these equations $Y_{lm}(\theta ,\phi )$ are the well--known spherical
harmonics, $j_{l}(z)$ is a spherical Bessel function and $x_{l,n}$ its nth
zero\cite{AS72}.

\section{Results and discussion}

\label{sec:results}

\noindent Table~\ref{tab:energies} shows the total, kinetic and potential
energies for the ground state for several values of the box radius $R$. The
first, second, and third rows for each entry shows perturbation and
variational results for the MNC and variational results for the CNC,
respectively. For simplicity, we restrict our MNC variational calculations
to the trial function
\begin{equation}
\varphi (r_{e},r_{n})=\left( 1-\frac{r_{e}}{R}\right) \left( 1-\frac{r_{n}}{R%
}\right) e^{-\alpha r_{e}-\beta r_{n}-\gamma r}\left(
c_{1}+c_{2}r+c_{3}r_{n}+c_{4}r_{e}\right)  \label{eq:phi_v_part}
\end{equation}
In the CNC we consider Eq.~(\ref{eq:phi_v_CN}) with $N=4$ that yields as
much as five--digits accuracy when compared with the higly accurate results
of Aquino et al\cite{ACM07}. On the other hand, the approximate variational
MNC energy is expected to be less accurate because the basis set in Eq.~(\ref
{eq:phi_v_part}) is comparatively smaller. However, present results are
considerably more accurate than previous ones for this model\cite{F10a,F10b}.

First--order perturbation theory is acceptable in the strong--coupling
region $R<1$ au. We appreciate that the perturbation result is
systematically greater than the variational one that is known to be a
rigorous upper bound. This fact clearly shows that the latter is more
accurate even for $R=0.1$ au where the perturbation result is expected to be
fairly accurate. Table~\ref{tab:energies} shows that perturbation theory
underestimates the nuclear kinetic energy and overestimates the potential
energy while, on the other hand, provides a quite reasonable estimate of the
electronic kinetic energy.

It is also interesting to compare the variational results for the MNC and
CNC. The MNC energy is greater than the CNC one for all values of $R$. The
reason is that the CNC energy is smallest when the nucleus is located at the
center of the box that is the particular case chosen here. This effect was
discussed earlier by means of a less accurate trial function for the MNC\cite
{F10a,F10b}. Table~\ref{tab:energies}  shows that both the kinetic and
potential energies are greater for the MNC.

Table~\ref{tab:var_par} shows the variational parameters for the trial
functions (\ref{eq:phi_v_part}) and (\ref{eq:phi_v_CN}) as well as the
expectation values of $r_{e}$, $r_{n}$ and $r$. Since the magnitude of the
nonlinear variational parameter $\alpha $ is negligible for all values of $R$
we have set it equal to zero. On the other hand, $\beta $ is quite large for
small $R$ and decreases as $R$ increases. The remaining nonlinear parameter $%
\gamma $ decreases with $R$ reaches a minimum and then increases
asymptotically towards the free--atom value. Those exponential parameters
suggest that the nucleus is localized about the box center whereas the
electron is localized about the nucleus (though not so strongly because $%
\beta >\gamma $ for all $R$). The fact that $\left\langle r_{n}\right\rangle
<\left\langle r_{e}\right\rangle \approx R/2$ supports this conjecture.
Table~\ref{tab:var_par} also shows that $\left\langle r\right\rangle
_{MNC}\approx \left\langle r\right\rangle _{CNC}$.

\vspace{4 mm}
{\Large
\centerline{\bf Acknowledgments}
}

\vspace{4 mm}
Two of us (NA and AFR) would like to thank financial support provided by {\em
Sistema Nacional de Investigadores} (SNI, Mexico).

\begin{table}[tbp]
\caption{Total, kinetic and potential energies for the confined hydrogen
atom for several values of the box radius. The entries stand for MNC
perturbation theory, MNC variational method and CNC variational method,
respectively}
\label{tab:energies}
\begin{center}
\begin{tabular}{c|c|cccc}
\hline
$R$ & $E$ & $\langle T \rangle$ & $\langle T_e \rangle$ & $\langle T_n
\rangle$ & $\langle V \rangle$ \\ \hline
0.1 & 475.88825 & 493.74898 & 493.48022 & 0.26876 & $-$17.86073 \\
& 473.84272 & 497.55784 & 495.52046 & 2.03739 & $-$23.71513 \\
& 468.99313 & 493.59225 & $-$ & $-$ & $-$24.59911 \\ \hline
0.2 & 114.50688 & 123.43724 & 123.37006 & 0.06719 & $-$8.93037 \\
& 112.47785 & 124.54287 & 123.91907 & 0.62381 & $-$12.06502 \\
& 111.07107 & 123.48629 & $-$ & $-$ & $-$12.41522 \\ \hline
0.3 & 48.90742 & 54.86100 & 54.83114 & 0.02986 & $-$5.95358 \\
& 47.28928 & 55.44914 & 55.13094 & 0.31820 & $-$8.15987 \\
& 46.59279 & 54.94902 & $-$ & $-$ & $-$8.35623 \\ \hline
0.4 & 26.39413 & 30.85931 & 30.84251 & 0.01680 & $-$4.46518 \\
& 25.06003 & 31.26099 & 31.06244 & 0.19855 & $-$6.20096 \\
& 24.63398 & 30.96278 & $-$ & $-$ & $-$6.32881 \\ \hline
0.5 & 16.17781 & 19.74996 & 19.73921 & 0.01075 & $-$3.57215 \\
& 15.03997 & 20.06358 & 19.92556 & 0.13802 & $-$5.02361 \\
& 14.74805 & 19.86217 & $-$ & $-$ & $-$5.11412 \\ \hline
0.6 & 10.73846 & 13.71525 & 13.70778 & 0.00747 & $-$2.97679 \\
& 9.74237 & 13.98064 & 13.87803 & 0.10261 & $-$4.23827 \\
& 9.52774 & 13.83361 & $-$ & $-$ & $-$4.30586 \\ \hline
0.7 & 7.52498 & 10.07651 & 10.07102 & 0.00548 & $-$2.55153 \\
& 6.63550 & 10.31303 & 10.23317 & 0.07986 & $-$3.67754 \\
& 6.46994 & 10.19984 & $-$ & $-$ & $-$3.72990 \\ \hline
0.8 & 5.48224 & 7.71483 & 7.71063 & 0.00420 & $-$2.23259 \\
& 4.67560 & 7.93312 & 7.86889 & 0.06424 & $-$3.25753 \\
& 4.54339 & 7.84251 & $-$ & $-$ & $-$3.29912 \\ \hline
0.9 & 4.11114 & 6.09567 & 6.09235 & 0.00332 & $-$1.98453 \\
& 3.37058 & 6.30211 & 6.24913 & 0.5298 & $-$2.93153 \\
& 3.26219 & 6.22741 & $-$ & $-$ & $-$2.96522 \\ \hline
1.0 & 3.15142 & 4.93749 & 4.93480 & 0.00269 & $-$1.78607 \\
& 2.46468 & 5.13617 & 5.09161 & 0.04456 & $-$2.67149 \\
& 2.37399 & 5.07314 & $-$ & $-$ & $-$2.69915 \\ \hline
1.1 & $-$ & $-$ & $-$ & $-$ & $-$ \\
& 1.81475 & 4.27426 & 4.23620 & 0.03807 & $-$2.45952 \\
& 1.73761 & 4.22006 & $-$ & $-$ & $-$2.48245 \\ \hline
1.2 & $-$ & $-$     & $-$ & $-$ & $-$ \\
& 1.33794 & 3.64094 & 3.61044 & 0.03049 & $-$2.30300 \\
& 1.26931 & 3.57213 & $-$ & $-$ & $-$2.30282 \\ \hline
1.3 & $-$ & $-$     & $-$ & $-$ & $-$ \\
& 0.97551 & 3.12685 & 3.10009 & 0.02676 & $-$2.15133 \\
& 0.91704 & 3.06876 & $-$ & $-$ & $-$2.15172 \\ \hline
1.4 & $-$ & $-$ & $-$ & $-$ & $-$ \\
& 0.69758 & 2.71961 & 2.69594 & 0.02367 & $-$2.02203 \\
& 0.64711 & 2.67018 & $-$ & $-$ & $-$2.02307 \\ \hline
1.5 & $-$ & $-$     & $-$ & $-$ & $-$ \\
& 0.48107 & 2.39175 & 2.37066 & 0.02110 & $-$1.91068 \\
& 0.43702 & 2.34943 & $-$ & $-$ & $-$1.91241 \\ \hline
\end{tabular}
\end{center}
\end{table}

\begin{table}[tbp]
\begin{center}
\begin{tabular}{c|c|cccc}
\hline
$R$ & $E$ & $\langle T \rangle$ & $\langle T_e \rangle$ & $\langle T_n
\rangle$ & $\langle V \rangle$ \\ \hline
2.0 & $-$    & $-$     & $-$     & $-$     & $-$ \\
& $-$0.09946 & 1.43084 & 1.41801 & 0.01283 & $-$1.53031 \\
& $-$0.12500 & 1.41016 & $-$ & $-$ & $-$1.53515 \\ \hline
3.0 & $-$    & $-$     & $-$ & $-$ & $-$ \\
& $-$0.41078 & 0.78067 & 0.77452 & 0.00615 & $-$1.19145 \\
& $-$0.42397 & 0.77206 & $-$ & $-$ & $-$1.19603 \\ \hline
4.0 & $-$    & $-$     & $-$ & $-$ & $-$ \\
& $-$0.47520 & 0.59178 & 0.58854 & 0.00325 & $-$1.06699 \\
& $-$0.48327 & 0.58486 & $-$ & $-$ & $-$1.06813 \\ \hline
5.0 & $-$    & $-$     & $-$ & $-$ & $-$ \\
& $-$0.49138 & 0.53155 & 0.52961 & 0.00194 & $-$1.02293 \\
& $-$0.49642 & 0.52453 & $-$ & $-$ & $-$1.02095 \\ \hline
6.0 & $-$    & $-$     & $-$ & $-$ & $-$ \\
& $-$0.49615 & 0.51194 & 0.51064 & 0.00130 & $-$1.00809 \\
& $-$0.49928 & 0.50635 & $-$ & $-$ & $-$1.00563 \\ \hline
7.0 & $-$ & $-$ & $-$ & $-$ & $-$ \\
& $-$0.49784 & 0.50515 & 0.50420 & 0.00096 & $-$1.00299 \\
& $-$0.49986 & 0.50148 & 0.50148 & $-$ & $-$1.00135 \\ \hline
8.0 & $-$ & $-$ & $-$ & $-$ & $-$ \\
& $-$0.49857 & 0.50256 & 0.50180 & 0.00075 & $-$1.00113 \\
& $-$0.49997 & 0.50032 & 0.50032 & $-$ & $-$1.00029 \\ \hline
9.0 & $-$ & $-$ & $-$ & $-$ & $-$ \\
& $-$0.49894 & 0.50145 & 0.50082 & 0.00062 & $-$1.00039 \\
& $-$0.50000 & 0.50006 & 0.50006 & $-$ & $-$1.00006 \\ \hline
10.0 & $-$ & $-$ & $-$ & $-$ & $-$ \\
& $-$0.49916 & 0.50070 & 0.50013 & 0.00057 & $-$0.99985 \\
& $-$0.50000 & 0.50002 & 0.50002 & $-$ & $-$1.00002 \\ \hline
\end{tabular}
\end{center}
\end{table}

\begin{table}[tbp]
\caption{Variational parameters for the trial functions (\ref{eq:phi_v_part}%
) and (\ref{eq:phi_v_CN}) ($N=4$) and some
expectation values}
\label{tab:var_par}
\begin{center}
\begin{tabular}{c|ccccccc|ccc}
\hline
$R$ & $\beta$ & $\gamma$ & $c_1$ & $c_2$ & $c_3$ & $c_4$ &  & $\langle r_e
\rangle$ & $\langle r_n \rangle$ & $\langle r \rangle_{MNC}$ \\
&  & $\delta$ & $d_0$ & $d_1$ & $d_2$ & $d_3$ & $d_4$ &  &  & $\langle r
\rangle_{CNC}$ \\ \hline
0.1 & 71.386 & 12.285 & 0.056 & 1.000 & $-$0.210 & 0.446 &  & 0.050 & 0.017
& 0.052 \\
& $-$ & 4.915 & 0.001 & 0.014 & 0.004 & $-$1.154 & 2.836 & $-$ & $-$ & 0.050
\\ \hline
0.2 & 40.494 & 6.305 & 0.120 & 1.000 & $-$0.203 & 0.514 &  & 0.099 & 0.031 &
0.102 \\
& $-$ & 1.068 & 0.095 & 0.531 & $-$2.000 & $-$1.603 & $-$0.639 & $-$ & $-$ &
0.099 \\ \hline
0.3 & 29.340 & 4.287 & 0.195 & 1.000 & $-$0.203 & 0.588 &  & 0.148 & 0.044 &
0.152 \\
& $-$ & 1.321 & 0.140 & 0.551 & $-$0.946 & $-$1.275 & $-$0.567 & $-$ & $-$ &
0.147 \\ \hline
0.4 & 23.400 & 3.267 & 0.286 & 1.000 & $-$0.207 & 0.670 &  & 0.196 & 0.055 &
0.200 \\
& $-$ & 1.267 & 0.261 & 0.767 & $-$0.768 & $-$1.127 & $-$0.513 & $-$ & $-$ &
0.195 \\ \hline
0.5 & 19.647 & 2.651 & 0.397 & 1.000 & $-$0.214 & 0.767 &  & 0.243 & 0.066 &
0.248 \\
& $-$ & 1.217 & 0.419 & 0.977 & $-$0.593 & $-$0.974 & $-$0.437 & $-$ & $-$ &
0.242 \\ \hline
0.6 & 17.031 & 2.237 & 0.535 & 1.000 & $-$0.225 & 0.883 &  & 0.290 & 0.077 &
0.295 \\
& $-$ & 1.156 & 0.617 & 1.171 & $-$0.446 & $-$0.848 & $-$0.333 & $-$ & $-$ &
0.289 \\ \hline
0.7 & 15.088 & 1.939 & 0.694 & 0.974 & $-$0.233 & 1.000 &  & 0.337 & 0.087 &
0.341 \\
& $-$ & 1.095 & 0.797 & 1.253 & $-$0.293 & $-$0.706 & $-$0.216 & $-$ & $-$ &
0.335 \\ \hline
0.8 & 13.578 & 1.713 & 0.781 & 0.825 & $-$0.215 & 1.000 &  & 0.382 & 0.097 &
0.387 \\
& $-$ & 1.041 & 0.958 & 1.262 & $-$0.163 & $-$0.586 & $-$0.126 & $-$ & $-$ &
0.380 \\ \hline
0.9 & 12.365 & 1.536 & 0.867 & 0.685 & $-$0.197 & 1.000 &  & 0.427 & 0.107 &
0.431 \\
& $-$ & 1.001 & 1.091 & 1.225 & $-$0.052 & $-$0.485 & $-$0.065 & $-$ & $-$ &
0.425 \\ \hline
1.0 & 11.365 & 1.393 & 0.954 & 0.551 & $-$0.182 & 1.000 &  & 0.471 & 0.117 &
0.475 \\
& $-$ & 1.011 & 0.972 & 0.980 & 0.073 & $-$0.330 & $-$0.033 & $-$ & $-$ &
0.468 \\ \hline
1.1 & 17.414 & 0.480 & 0.025 & $-$0.091 & 1.000 & 0.095 &  & 0.515 & 0.123 &
0.518 \\
& $-$ & 1.212 & 0.885 & 1.000 & 0.233 & $-$0.074 & $-$0.124 & $-$ & $-$ &
0.511 \\ \hline
1.2 & 16.215 & 0.481 & 0.028 & $-$0.095 & 1.000 & 0.099 &  & 0.557 & 0.132 &
0.560 \\
& $-$ & 1.216 & 0.948 & 1.000 & 0.272 & $-$0.022 & $-$0.099 & $-$ & $-$ &
0.553 \\ \hline
1.3 & 15.168 & 0.482 & 0.031 & $-$0.100 & 1.000 & 0.103 &  & 0.599 & 0.141 &
0.602 \\
& $-$ & 1.224 & 0.993 & 0.989 & 0.298 & 0.019 & $-$0.076 & $-$ & $-$ & 0.595
\\ \hline
1.4 & 14.243 & 0.485 & 0.033 & $-$0.104 & 1.000 & 0.107 &  & 0.640 & 0.150 &
0.642 \\
& $-$ & 1.264 & 1.000 & 0.983 & 0.328 & 0.071 & $-$0.061 & $-$ & $-$ & 0.635
\\ \hline
1.5 & 13.418 & 0.488 & 0.037 & $-$0.109 & 1.000 & 0.112 &  & 0.681 & 0.159 &
0.681 \\
& $-$ & 1.279 & 1.000 & 0.951 & 0.335 & 0.093 & $-$0.041 & $-$ & $-$ & 0.675
\\ \hline
\end{tabular}
\end{center}
\end{table}

\begin{table}[tbp]
\begin{center}
\begin{tabular}{c|ccccccc|ccc}
\hline
$R$ & $\beta$ & $\gamma$ & $c_1$ & $c_2$ & $c_3$ & $c_4$ &  & $\langle r_e
\rangle$ & $\langle r_n \rangle$ & $\langle r \rangle_{MNC}$ \\
&  & $\delta$ & $d_0$ & $d_1$ & $d_2$ & $d_3$ & $d_4$ &  &  & $\langle r
\rangle_{CNC}$ \\ \hline
2.0 & 10.303 & 0.510 & 0.056 & $-$0.131 & 1.000 & 0.133 &  & 0.869 & 0.204 &
0.864 \\
& $-$ & 1.348 & 1.000 & 0.841 & 0.385 & 0.062 & 0.045 & $-$ & $-$ & 0.859 \\
\hline
3.0 & 4.498 & 0.565 & 1.000 & $-$0.459 & 0.312 & 0.452 &  & 1.178 & 0.319 &
1.147 \\
& $-$ & 0.922 & 1.000 & 0.243 & 0.091 & $-$0.012 & 0.012 & $-$ & $-$ & 1.153
\\ \hline
4.0 & 2.974 & 0.642 & 1.000 & $-$0.324 & $-$0.015 & 0.319 &  & 1.391 & 0.449
& 1.320 \\
& $-$ & 0.400 & 1.000 & $-$0.350 & 0.086 & $-$0.012 & 0.001 & $-$ & $-$ &
1.342 \\ \hline
5.0 & 2.159 & 0.711 & 1.000 & $-$0.254 & $-$0.063 & 0.251 &  & 1.539 & 0.599
& 1.408 \\
& $-$ & 0.424 & 1.000 & $-$0.375 & 0.087 & $-$0.012 & 0.001 & $-$ & $-$ &
1.440 \\ \hline
6.0 & 1.643 & 0.768 & 1.000 & $-$0.200 & $-$0.090 & 0.201 &  & 1.657 & 0.761
& 1.450 \\
& $-$ & 0.465 & 1.000 & $-$0.367 & 0.079 & $-$0.010 & 0.001 & $-$ & $-$ &
1.481 \\ \hline
7.0 & 1.296 & 0.813i & 1.000 & $-$0.157 & $-$0.109 & 0.165 &  & 1.770 & 0.930
& 1.470 \\
& $-$ & 0.510 & 1.000 & $-$0.346 & 0.068 & $-$0.008 & 0.000 & $-$ & $-$ &
1.495 \\ \hline
8.0 & 1.043 & 0.851 & 1.000 & $-$0.122 & $-$0.126 & 0.139 &  & 1.887 & 1.106
& 1.480 \\
& $-$ & 0.555 & 1.000 & $-$0.319 & 0.057 & $-$0.006 & 0.000 & $-$ & $-$ &
1.499 \\ \hline
9.0 & 0.841 & 0.886 & 1.000 & $-$0.089 & $-$0.142 & 0.118 &  & 2.018 & 1.295
& 1.486 \\
& $-$ & 0.601 & 1.000 & $-$0.287 & 0.046 & $-$0.004 & 0.000 & $-$ & $-$ &
1.500 \\ \hline
10.0 & 0.747 & 0.910 & 1.000 & $-$0.067 & $-$0.151 & 0.102 &  & 2.104 & 1.412
& 1.490 \\
& $-$ & 0.935 & 1.000 & 0.032 & 0.009 & $-$0.001 & 0.000 & $-$ & $-$ & 1.500
\\ \hline
\end{tabular}
\par
\end{center}
\end{table}


\begin{thebibliography}{99}
\bibitem{F84}  P. W. Fowler, Molec. Phys. \textbf{53}, 865 (1984).

\bibitem{PV09}  S. H. Patil and Y. P. Varshni, Adv. Quantum Chem. \textbf{57}%
, 1 (2009).

\bibitem{SPM09}  K. D. Sen, V. I. Pupyshev, and H. E. Montgomery Jr, Adv.
Quantum Chem. \textbf{57}, 25 (2009).

\bibitem{L-K09}  E. Ley-Koo, Adv. Quantum Chem. \textbf{57}, 79 (2009).

\bibitem{A09}  N. Aquino, Adv. Quantum Chem. \textbf{57}, 123 (2009).

\bibitem{L09}  C. Laughlin, Adv. Quantum Chem. \textbf{57}, 203 (2009).

\bibitem{MRU95}  J. L. Mar\'{i}n, R. Rosas, and A. Uribe, Am. J. Phys.
\textbf{63}, 460 (1995).

\bibitem{F10a}  F. M. Fern\'{a}ndez, Eur. J. Phys. \textbf{31}, 285 (2010).

\bibitem{F10b}  F. M. Fern\'{a}ndez, Eur. J. Phys. \textbf{31}, 611 (2010).

\bibitem{F10c}  F. M. Fern\'{a}ndez, Perturbation theory for confined
systems, arXiv:1004.2508v1 [quant-ph]

\bibitem{FC87}  F. M. Fern\'{a}ndez and E. A. Castro, Hypervirial theorems (
Springer, Berlin, Heidelberg, New York, London, Paris, Tokyo, 1987).

\bibitem{H64}  E. A. Hylleraas, Adv. Quantum Chem. \textbf{1}, 1 (1964).

\bibitem{KRS60}  W. Kolos, C. C. J. Roothaan, and R. A. Sack, Rev. Mod.
Phys. \textbf{32}, 178 (1960).

\bibitem{SMKS87}  K. Szalewicz, H. J. Monkhorst, W. Kolos, and A. Scrinzi,
Phys. Rev. A \textbf{36}, 5494 (1987).

\bibitem{FF93}  P. Froelich and A. Flores-Riveros, Phys. Rev. Lett. \textbf{%
70}, 1595 (1993).

\bibitem{AS72}  M. Abramowitz and I. A. Stegun, Handbook of Mathematical
Functions, Ninth ed. (Dover, New York, 1972).

\bibitem{ACM07}  N. Aquino, G. Campoy, and H. E. Montgomery Jr, Int. J.
Quantum Chem. \textbf{107}, 1548 (2007).
\end{thebibliography}
\end{document}